\documentclass[
superscriptaddress,dvips
prb,twocolumn
]{revtex4-1}

\usepackage{graphicx,amsmath,amssymb,subfigure}
\usepackage{dcolumn}
\usepackage{bm}
\usepackage{color}

\begin{document}

\preprint{APS/123-QED}

\title{Spin transport in the N\'{e}el and collinear antiferromagnetic phase of the two dimensional spatial and spin anisotropic Heisenberg model on a square lattice}
\author{Zewei Chen}
\affiliation{State Key Laboratory of Optoelectronic Materials and Technologies,\\
School of Physics and Engineering, Sun Yat-sen University, Guangzhou 510275, China}
\author{Trinanjan Datta}
\email[Corresponding author:]{tdatta@aug.edu}
\affiliation{Department of Chemistry and Physics, Augusta State University, Augusta, GA 30904}
\affiliation{State Key Laboratory of Optoelectronic Materials and Technologies,\\
School of Physics and Engineering, Sun Yat-sen University, Guangzhou 510275, China}
\author{Dao-Xin Yao}
\email[Corresponding author:]{yaodaox@mail.sysu.edu.cn}
\affiliation{State Key Laboratory of Optoelectronic Materials and Technologies,\\
School of Physics and Engineering, Sun Yat-sen University, Guangzhou 510275, China}

\date{\today}

\begin{abstract}
We analyze and compare the effect of spatial and spin anisotropy on spin conductivity in a two dimensional S=1/2 Heisenberg quantum magnet on a square lattice. We explore the model in both the N\'{e}el antiferromagnetic (AF) phase and the collinear antiferromagnetic (CAF) phase. We find that in contrast to the effects of spin anisotropy ($\Delta$) in the Heisenberg model, spatial anisotropy ($\eta$) in the AF phase does not suppress the zero temperature regular part of the spin conductivity, $\sigma^{reg}(\omega)$, in the zero frequency limit - rather it enhances it.  We also explore the finite temperature (T) effects on the Drude weight, D$_{S}(\eta, \Delta, T)$, in the AF phase for various spatial and spin anisotropy parameters. We find that D$_{S}$ goes to zero as the temperature approaches zero. At finite temperatures (within the collision less approximation) enhancing spatial anisotropy increases the Drude weight value and increasing spin anisotropy decreases the Drude weight value. In the CAF phase (within the non-interacting approximation) the zero frequency spin conductivity has a finite value for non-zero values of the spatial anisotropy parameter. In the CAF phase increasing the spatial anisotropy parameter suppresses the $\sigma^{reg}(\omega)$ response at zero frequency. Furthermore, we find that the CAF phase displays a spike in the spin conductivity not seen in the AF phase. Inclusion of the smallest amount of spin anisotropy causes $\sigma^{reg}(\omega)$ to develop a gap in the spin conductivity response of both the AF and CAF phase. Based on these studies \emph{we conclude that materials with spatial anisotropy are better spin conductors than those with spin anisotropy both at zero and finite temperatures}. We utilize exchange parameter ratios for real material systems such as SrZnVO(PO$_{4}$)$_{2}$ (spatially anisotropic) and La$_{2}$NiO$_{4}$ (spin isotropic) as inputs to the computation of spin conductivity.
\begin{description}
\item[PACS numbers] 75.10.Jm, 75.30.Ds, 75.40.Gb, 75.76.+j
\end{description}
\end{abstract}
\maketitle

\section{Introduction}
The concept of spin transport is both intriguing and exciting ~\cite{pulizzinatmat368,maekawa}. The importance of studying spin transport is related to the field of spintronics which involves the study of active control and manipulation of spin degrees of freedom in a solid state material ~\cite{wolfscience16112001,slonPRB396995,sarmaRMP76323,martinPRL87187202}. The fundamental quantity of interest is spin conductivity associated with the flow of spin current. Charge currents in two-dimensional high-mobility electron systems with Rashba spin-orbit coupling can generate spin currents ~\cite{bychkovJPC176039,murakami05092003,sinovaPRL92126603}. However, power dissipation in spin systems in which spin transport is accompanied by charge transport limits the possible application to state-of-the-art spintronic devices. In nonitinerant quantum systems the dissipation problem is reduced. This is a major technological motivation behind exploring the fundamental physical processes of spin transport in insulating magnets ~\cite{lossPRL101017202}.  

The definition of spin current is a controversial topic ~\cite{schutzepjb557,shiPRL96076604,soninAP59181,tokatlyPRL101106601}. The issue arises in the context of spin transport in semiconductors where spin is not conserved because of the presence of spin-orbit interaction. For insulating magnets the spin current operator definition, $\mathcal{{\bf J}}_{i \rightarrow j}$, has been investigated by Sch\"{u}tz et ~al. It has been shown that for a collinear spin configuration (the case considered in this paper) magnetization transport is appropriately described by the scalar current density operator defined in Eq.~\ref{eq:jxop}. Therefore issues associated with the conceptual definition of spin current are of no concern for the spin conductivity analysis carried out in this paper. Possible experimental set up to measure spin conductivity have been proposed in Ref.~\onlinecite{lossPRL90167204}.

Spin transport phenomena has been explored in ferromagnetic~\cite{maekawanat8777,maekawaPRB83094410} and  antiferromagnetic (AF) insulators ~\cite{zotosPRB53983,piresPRB79064401,piresJPC21245502,meierPRL90167204,lossPRB85054413,lossPRB84024402,wangPRB73212413}. Low-dimensional magnetic systems have been investigated using the spin anisotropic Heisenberg AF model for both S=1/2 and S=1. The 2D \& 3D Heisenberg model ~\cite{sentefPRB75214403} and 2D XY magnets ~\cite{limaEPJB101140} have also been analyzed for spin transport. Additional work has been done on the effects of Dzyaloshinskii-Moriya interactions on spin Heisenberg magnets in one- and two- dimensions ~\cite{limaPSSB201147167}. Present analysis on the models explored till date show unconventional ballistic spin transport at finite temperatures. For the two-dimensional (2D) AF the regular part of the spin conductivity remains finite in the dc limit (for the isotropic Heisenberg system) ~\cite{sentefPRB75214403}, at zero temperature, while for the 1D case the regular part of the conductivity is suppressed at low frequencies~\cite{piresPRB79064401,piresJPC21245502}. Two-magnon processes contribute to the magnetization transport and can be attributed to the motion of quasi-particles as magnons (2D) or spinons (1D). 

\begin{figure}[t]
\centering
\includegraphics[width=3.0in]{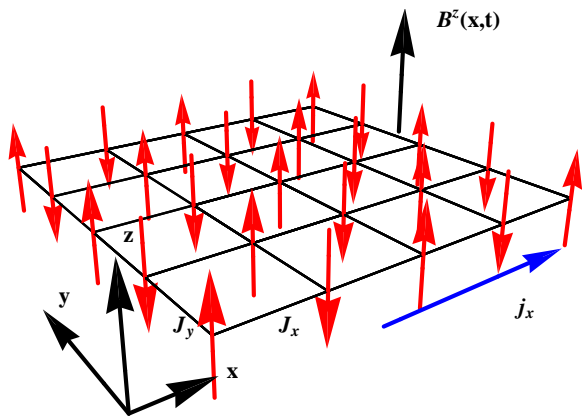}
\caption{\label{fig:anisheis} (Color online) The spatial and spin anisotropic Heisenberg model considered in this paper (see Eq.~\ref{eq:ham}). Exchange interactions along the x- and y- direction are given by J$_{x}$ and J$_{y}$ respectively. The z- component spin anisotropy is given by $\Delta$. The above image is for the model in the antiferromagnetic phase. The model is studied both in the N\'{e}el antiferromagnetic and collinear antiferromagnetic phase. The external magnetic field B$^{z}$(x,t) is directed along the z- direction and is spatially inhomogeneous along the x- direction. The inhomogeneous magnetic field drives a spin current, j$_{x}$, along the x- direction. In this article we explore the effects of spatial and spin anisotropy on the spin current j$_{x}$.} 
\end{figure}
   
In this paper we study the effects of spin conductivity on a spatial and spin anisotropic S = 1/2 Heisenberg model (see Fig.~\ref{fig:anisheis}). The model is not purely of academic interest since there are several vanadium phosphate material systems (Pb$_{2}$VO(PO$_{4}$)$_{2}$, SrZnVO(PO$_{4}$)$_{2}$, BaZnVO(PO$_{4}$)$_{2}$, and BaCdVO(PO$_{4}$)$_{2}$) in which extensive bandstructure calculations show a spatially anisotropic exchange interaction along the x- and y- direction ~\cite{alexPRB79214417}. For isotropic exchange interactions we consider La$_{2}$NiO$_{4}$ as the model system~\cite{yaoPRB80144523}. We utilize these real material exchange parameter ratios as inputs to compute the spin conductivity response in the N\'{e}el AF phase. We discuss the temperature (T) dependence of the spin Drude weight, D$_{S}(\eta,\Delta,T)$, on spatial and spin anisotropy within the collision less approximation scheme for the magnons. We also analyze the spin conductivity behavior of the previously unexplored collinear antiferromagnetic (CAF) phase. 

\begin{figure}[b]
\centering
\includegraphics[width=3.5in]{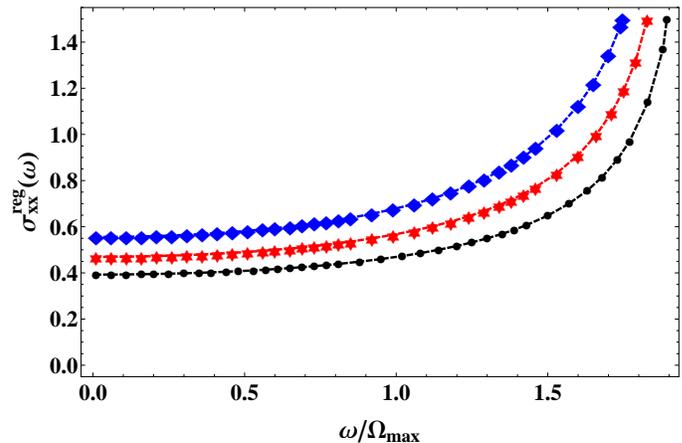}
\caption{\label{fig:sigmani} (Color online) Regular part of the non-interacting spin conductivity, $\sigma^{reg}_{xx}(\omega)$, in units of (g$\mu_{B})^{2}$/h for the spatially anisotropic but spin isotropic ($\Delta=1$) 2D Heisenberg model in the antiferromagnetic phase. Here $\widetilde{\omega}=\omega/\Omega_{max}$ where $\Omega_{max}=2J_{x}(1+\eta)S\alpha(S)/\hbar$. The plot is for $\eta=J_{y}/J_{x}$ values of 1 (black circles), 0.7 (red stars), and 0.5 (blue diamonds). The dashed lines are a guide to the eye. Decreasing $\eta$ leads to an increase in the regular part of the spin conductivity. The divergence near $\widetilde{\omega}$=2 is due to the Van-Hove singularity.}
\end{figure}

\begin{figure}[h]
\includegraphics[width=3.5in]{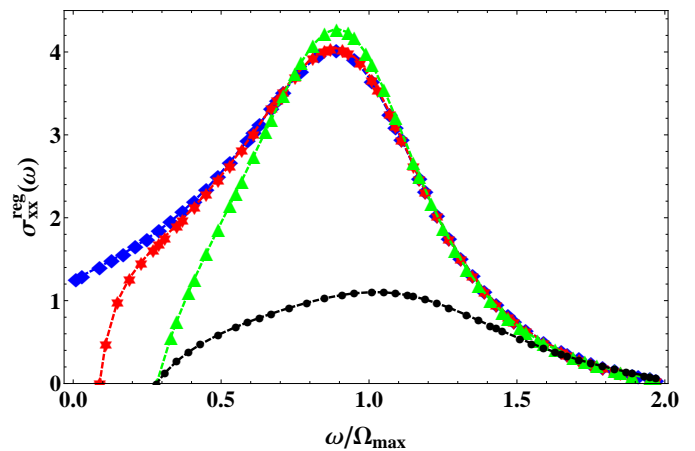}
\caption{\label{fig:sigmaietadelta} (Color online) Regular part of the interacting spin conductivity, $\sigma^{reg}_{xx}(\omega)$, in units of (g$\mu_{B})^{2}$/h within ladder approximation for the spatial and spin anisotropic model.  Here $\widetilde{\omega}=\omega/\Omega_{max}$ where $\Omega_{max}=2J_{x}\Delta(1+\eta)S\alpha(S)/\hbar$. The plot is for $\eta=0.5~\& ~\Delta=1$ (blue diamonds), $\eta=0.5  ~\& ~\Delta=1.001$ (red stars), $\eta=0.5  ~\& ~\Delta=1.01$ (green triangles), and $\eta=1.5  ~\& ~\Delta=1.01$ (black circles). The dashed lines are a guide to the eye.}
\end{figure}

\begin{figure}[t]
\centering
\includegraphics[width=3.5in]{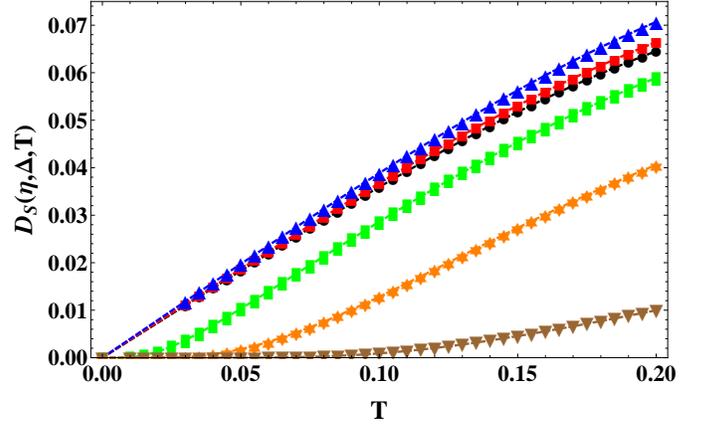}
\caption{\label{fig:neeldrude} (Color online) Temperature (T), spatial anisotropy parameter ($\eta$), and spin anisotropy parameter ($\Delta$) dependence of the Drude weight, D$_{S}$(T, $\eta$, $\Delta$). The plot is for $\eta=1~\&~\Delta=1$ (black circles), $\eta=0.7~\& ~\Delta=1$ (red squares), $\eta=0.5~\&~\Delta=1$ (blue up triangles), $\eta=1~\& ~\Delta=1.001$ (green rectangles), $\eta=1~\&~\Delta=1.01$ (orange stars), and $\eta=1~\&~\Delta=1.1$ (brown down triangles). The dashed lines are a guide to the eye.}
\end{figure}

The main results of this paper can be summarized as follows (i) In contrast to the spin anisotropic Heisenberg model, spatial anisotropy in the AF phase does not suppress the regular part of the spin conductivity in the zero frequency limit. Rather it helps to enhance the spin conductivity. Compared to spatial anisotropy we find that spin anisotropy strongly suppresses the zero frequency weight of the regular part of the spin conductivity (see Figs.~\ref{fig:sigmani}, \ref{fig:sigmaieta}, and \ref{fig:sigmaietadelta}), (ii) as expected the finite temperature Drude weight in the AF phase vanishes in the zero temperature limit (this is the same behavior displayed in the spin anisotropic 2D Heisenberg model). Increasing spatial anisotropy increases the Drude weight value. However, increasing the spin anisotropy parameter decreases the Drude weight value (see Fig.~\ref{fig:neeldrude}). The Drude weight results are obtained within the non-interacting (collision less) magnon approximation (see Section ~\ref{subsec:drude} for discussion) and (iii) spin conductivity in the CAF phase has a non vanishing zero frequency contribution. In contrast to the AF phase, spatial anisotropy in the CAF phase decreases the zero frequency regular spin conductivity value. The CAF phase also displays a spike in the spin conductivity. Based on the above analysis we conclude that materials with spatial anisotropy are better spin conductors than those with spin anisotropy in both the AF and CAF phase at zero and finite temperatures.

This paper is organized as follows. In Section~\ref{sec:hamiltonian} we introduce the model Hamiltonian for the AF phase and derive the Fourier transformed Dyson-Maleev (DM) version of the Hamiltonian upto quartic interactions. The vertices for the quartic interaction are listed in Appendix~\ref{app:appvert}. In Section~\ref{sec:spincurrent} we state and derive the expression for the spin current operator for our model. In Section~\ref{sec:spincon} we utilize the many body Green function formalism to compute the spin conductivity. In Section~\ref{subsec:scaf} we  present and discuss for the AF phase the non-interacting (Section~\ref{subsubsec:nispincon}) and interacting spin conductivity (Section~\ref{subsubsec:ladder}) results. In Section~\ref{subsec:drude} we discuss the anisotropy and temperature dependence results of the drude weight contribution. In Section~\ref{subsec:sccaf} we introduce the model Hamiltonian for the CAF phase and present the results of the spin conductivity within the non-interacting approximation. Finally, in Section~\ref{sec:sum} we summarize the main findings of our work.

{\section{Model Hamiltonian}\label{sec:hamiltonian}}
The Hamiltonian for the 2D Heisenberg model on the square lattice with both directional anisotropy of exchange couplings and spin anisotropy is defined as
\begin{eqnarray}
H&=&\sum_{i,{\bf x}}J_{x}[\frac{1}{2}(S_{i}^{+}S_{i+{\bf x}}^{-}+S_{i}^{-}S_{i+{\bf x}}^{+})+\Delta S_{i}^{z}S_{i+{\bf x}}^{z}]\nonumber\\&+&
\sum_{i,{\bf y}}J_{y}[\frac{1}{2}(S_{i}^{+}S_{i+{\bf y}}^{-}+S_{i}^{-}S_{i+{\bf y}}^{+})+\Delta S_{i}^{z}S_{i+{\bf y}}^{z}],
\label{eq:ham}
\end{eqnarray}
The exchange interaction in the x-direction is denoted by J$_{x}$ and along the y-direction by J$_{y}$. The spin vector, {\bf S}$_{i}$, is defined on every lattice site where the index i runs over all lattice sites and i+{\bf x} and i +{\bf y} runs over nearest neighbors in the x- and y- direction respectively. S$^{+}_{i}$ and S$^{-}_{i}$ represent the raising and lowering spin operators at site i. S$^{z}_{i}$ represents the z-th component of the spin vector {\bf S}$_{i}$. The spin anisotropy parameter is given by $\Delta$ and the spatial anisotropy parameter by $\eta=\frac{J_{y}}{J_{x}}$. The magnetic model considered here supports three types of classical ground-state configurations - the N\'{e}el ($\pi$,$\pi$) AF state and columnar ($\pi$,0) or row (0,$\pi$) CAF phase. The AF state survives for $\eta > 0$ \& $\Delta \geq 1$. The CAF phase for $\eta < 0$ \& $\Delta \geq 1$. In Table~\ref{tab:FCCDatatable} we list examples of material systems with exchange interaction ratios and spin anisotropies of interest in this paper. 

We study the above Hamiltonian using the DM transformation. In the DM representation the spin operators are replaced by the bosonic operators as
\begin{eqnarray}
S^{z}_{i}=S-a^{\dag}_{i}a_{i},S^{-}_{i}=\sqrt{2S}a^{\dag}_{i},\nonumber\\
S^{+}_{i}=\sqrt{2S}\left(1-\frac{a^{\dag}_{i}a_{i}}{2S}\right)a_{i},
\label{eq:DMa}
\end{eqnarray}
for the up-spin (A) sublattice and by
\begin{eqnarray}
S^{z}_{j}=-S+b^{\dag}_{j}b_{j},S^{-}_{j}=\sqrt{2S}b_{j},\nonumber\\
S^{+}_{j}=\sqrt{2S}b^{\dag}_{j}\left(1-\frac{b^{\dag}_{j}b_{j}}{2S}\right),
\label{eq:DMb}
\end{eqnarray}
for the down-spin (B) sublattice. a$^{\dag}_{i}$(a$_{i}$) and b$^{\dag}_{i}$(b$_{i}$) represent the creation (annihiliation) operators on site i in the A or B sublattice respectively. The Fourier transformations for each operator a$_{i}$ and b$_{i}$ is defined below
\begin{equation}
a_{i}=\sqrt{\frac{2}{N}}\sum_{{\bf k}}e^{-i{\bf k}\cdot{\bf R}_{i}}a_{{\bf k}}, 
b_{j}=\sqrt{\frac{2}{N}}\sum_{{\bf k}}e^{+i{\bf k}\cdot{\bf R}_{j}}b_{{\bf k}},
\label{eq:ftrans} 
\end{equation}
where N is the number of lattice sites and the momentum {\bf k} is restricted to the magnetic Brillouin zone. Utilizing the definition of the Fourier transformations, the DM transformation in Eqs.~\ref{eq:DMa} and \ref{eq:DMb}, and the Bogoliubov transformations 
\begin{eqnarray}
a_{{\bf k}}=u_{{\bf k}}\alpha_{{\bf k}}+v_{{\bf k}}\beta^{\dag}_{{\bf k}},
b_{{\bf k}}=u_{{\bf k}}\beta_{{\bf k}}+v_{{\bf k}}\alpha^{\dag}_{{\bf k}},
\label{eq:bogo}
\end{eqnarray}
we can diagonalize the quadratic part of the Hamiltonian. The original Hamiltonian, Eq.~\ref{eq:ham}, can then be written in {\bf k}-space as
\begin{equation}
H_{DM}=E_{0}+H_{1}+:V_{DM}:
\label{eq:basichdm}
\end{equation}
where H$_{DM}$ is the DM transformed Hamiltonian. The classical energy (E$_{0}$) and the linear spin wave theory (LSWT) hamiltonian (H$_{1}$) is given by
\begin{eqnarray}
E_{0}&=&-N\Delta J_{x}(1+\eta)S^{2}\alpha^{2}(S),\\
H_{1}&=&\sum_{\bf{k}}\hbar\Omega_{\bf{k}}(\alpha_{\bf{k}}^{+}\alpha_{\bf{k}}+\beta_{\bf{k}}^{+}\beta_{\bf{k}}).
\label{eq:lswt}
\end{eqnarray}The coefficients u$_{{\bf k}}$ and v$_{{\bf k}}$ are given by,
\begin{equation}
u_{{\bf k}}=\sqrt\frac{1+\varepsilon_{{\bf k}}}{2\varepsilon_{{\bf k}}}, v_{{\bf k}}=-sgn(\gamma_{{\bf k}})\sqrt\frac{1-\varepsilon_{{\bf k}}}{2\varepsilon_{{\bf k}}}.
\label{eq:uv}
\end{equation}
The LSWT dispersion ($\Omega_{{\bf k}}$) and the defintion of $\gamma_{{\bf k}}$ is
\begin{equation}
\hbar \Omega_{\bf{k}}=2\Delta J_{x}(1+\eta)S\alpha(S)\varepsilon_{\bf{k}},
\label{eq:Omega}
\end{equation}
\begin{equation}
\varepsilon_{\bf{k}}=\sqrt{1-\gamma_{\bf{k}}^{2}/\Delta^{2}},
\label{eq:epsil}
\end{equation}
\begin{equation}
\gamma_{\bf{k}}=\frac{1}{(1+\eta)}(\cos k_{x}+\eta \cos k_{y}).
\label{eq:gammak}
\end{equation}
The Oguchi correction~\cite{oguchiPRev117117} $\alpha(S)$ is given by
\begin{equation}
\alpha(S)= 1+\frac{r}{2S};r=1 - \frac{2}{N}\sum_{{\bf k}}\varepsilon_{{\bf k}}.
\label{eq:oguchi}
\end{equation}
The normal ordered DM quartic interactions, :V$_{DM}$:, can be expressed in terms of the $\alpha$ and $\beta$ bosons as
\begin{widetext}
\begin{eqnarray}
:V_{DM}:&=&-\frac{J_{x}(1+\eta)}{N}\sum_{(1234)}\delta_{\bf{G}}(1+2-3-4)(V^{(1)}\alpha_{1}^{+}\alpha_{2}^{+}\alpha_{3}\alpha_{4}+V^{(2)}\alpha_{1}^{+}\beta_{2}\alpha_{3}\alpha_{4}
+V^{(3)}\alpha_{1}^{+}\alpha_{2}^{+}\beta_{3}^{+}\alpha_{4}+V^{(4)}\alpha_{1}^{+}\alpha_{3}\beta_{4}^{+}\beta_{2}
\nonumber\\&+&V^{(5)}\beta_{4}^{+}\alpha_{3}\beta_{2}\beta_{1}+V^{(6)}\beta_{4}^{+}\beta_{3}^{+}\alpha_{2}^{+}\beta_{1}+V^{(7)}\alpha_{1}^{+}\alpha_{2}^{+}\beta_{3}^{+}\beta_{4}^{+}+V^{(8)}\beta_{1}\beta_{2}\alpha_{3}\alpha_{4}
+V^{(9)}\beta_{4}^{+}\beta_{3}^{+}\beta_{2}\beta_{1}).
\label{eq:vdm}
\end{eqnarray}
\end{widetext}
where ${\bf G}$ is a reciprocal-lattice vector and the symbols 1, 2, 3, and 4 stand for wavevectors {\bf k}$_{1}$, {\bf k}$_{2}$, {\bf k}$_{3}$, and {\bf k}$_{4}$ respectively. The expression for the vertex functions, V$^{(m)}$, with m=1,2,...,9 are stated in Appendix A. The interaction vertices include the effect of spin anisotropy through the $\Delta$ parameter and is a generalization of the vertices stated in Ref.~\onlinecite{canaliPRB457127}. 
\begin{table}[t]
\caption{\label{tab:FCCDatatable} Exchange anisotropy parameter ratios used in the computation of spin conductivity. The spatial anisotropy parameter is given by $\eta$ and the spin anisotropy parameter by $\Delta$. See Figs.~\ref{fig:sigmani} - \ref{fig:cafd} for results.}
\begin{ruledtabular}
\begin{tabular}{ccc}
   Material                                &Type                          &$\eta$=J$_{y}$/J$_{x}$,$\Delta=1$ \\ \hline
   La$_{2}$NiO$_{4}$\tablenotemark[1]             &Spatially isotropic      &1, 1 \\
                                                &Spin isotropic             & \\
   SrZnVO(PO$_{4}$)$_{2}$\tablenotemark[2]   &Spatially anisotropic &0.7, 1 \\
                                                &Spin isotropic             &\\
    \end{tabular}
\end{ruledtabular}
\tablenotetext[1]{Ref.~\cite{yaoPRB80144523}}
\tablenotetext[2]{Ref.~\cite{alexPRB79214417}}
\end{table}
{\section{Spin current operator}\label{sec:spincurrent}}
It has been shown that the presence of an inhomogeneous magnetic field can drive a spin current ~\cite{sentefPRB75214403}. We consider a time dependent magnetic field directed along the z-direction with a spatial modulation along the x-axis, {\bf B}=B$^{z}(x,t)${{\bf\^{ z}}} (see Fig.~\ref{fig:anisheis}). The Hamiltonian, Eq.~\ref{eq:ham}, in the presence of time dependent external magnetic field can be rewritten as
\begin{equation}
H(t)=H-\sum_{x}S^{z}(x)B^{z}(x,t).
\end{equation}
Using the continuity equation for the S$^{z}$ component of the spin we can write down the basic definition of the spin current operator ~\cite{sentefPRB75214403}
\begin{equation}
{\bf j}_{i \rightarrow j}=\frac{i}{2}J_{ij}(S^{+}_{i}S^{-}_{j}-S^{-}_{i}S^{+}_{j}).
\label{eq:jxop}
\end{equation}
Since there is no symmetry breaking field present along the y-direction we do not expect any currents to flow in that direction. We therefore focus on the longitudinal spin transport current j$_{x}$ of the S$^{z}$ component of the magnetization. Using Eq.~\ref{eq:jxop} we obtain the following expression for the spin current operator in real space (within LSWT)
\begin{equation}
j_{x0}=SJi\left[\sum_{l\epsilon A}(a_{l}b_{l+x}-a_{l}^{+}b_{l+x}^{+})+\sum_{l\epsilon B}(b_{l}^{+}a_{l+x}^{+}-b_{l}a_{l+x})\right].
\label{eq:opersc}
\end{equation}
We then Fourier transform the spin current operator utilizing Eq.~\ref{eq:ftrans} to obtain
\begin{eqnarray}
j_{x0}=2J_{x}S\alpha(S)\sum_{\bf{k}}\sin(k_{x})\left[\frac{\gamma_{\bf{k}}}{\varepsilon_{\bf{k}}}(\alpha_{\bf{k}}^{+}\alpha_{\bf{k}}+\beta_{\bf{k}}^{+}\beta_{\bf{k}})\right]\nonumber\\
-2J_{x}S\alpha(S)\sum_{\bf{k}}\sin(k_{x})\left[\frac{1}{\varepsilon_{\bf{k}}}(\alpha_{\bf{k}}^{+}\beta_{\bf{k}}^{+}+\alpha_{\bf{k}}\beta_{\bf{k}})\right].
\label{eq:scft}
\end{eqnarray}
Spin conductivity in the long wavelength limit can be computed using the general definition provided in Ref.~\onlinecite{sentefPRB75214403}. We have
\begin{equation}
\sigma_{xx}=-(g\mu_{B})^{2}\frac{\langle -K_{x}\rangle -\Lambda_{xx}({\bf q}=0,\omega)}{i(\omega+i0^{+})},
\label{eq:spincon}
\end{equation}
where $\Lambda_{xx}$ is the longitudinal spin current correlation function
\begin{equation}
\Lambda_{xx}=\frac{i}{\hbar N}\int^{\infty}_{0}dt e^{i(\omega+i 0^{+})}\langle[j_{x}({\bf q},t),j_{x}(-{\bf q},0)]\rangle,
\end{equation}
and $\langle K_{x}\rangle$ is the spin-flip operator
\begin{equation}
\langle K_{x}\rangle=\frac{1}{2\hbar N}\sum_{i}\langle S^{+}_{i}S^{-}_{i+x}+S^{-}_{i}S^{+}_{i+x}\rangle.
\end{equation}
g is the gyromagnetic ratio, $\mu_{B}$ is the Bohr magneton, and $\hbar$ is the reduced Planck constant. Based on the above definitions the real part of the spin conductivity can be written as
\begin{equation}
\Re e[\sigma_{xx}] = D_{S}\delta(\omega)+\sigma^{reg}(\omega),
\end{equation}
where the spin Drude weight, D$_{S}$, associated with the singular part of the conductivity is defined as
\begin{equation}
\frac{D_{S}}{\pi}=(g\mu_{B})^{2}\{\langle -K_{x}\rangle -\Re e[\Lambda_{xx}({\bf q}=0,\omega\rightarrow 0)]\},
\label{eq:dsweight}
\end{equation}
and the regular part, $\sigma^{reg}_{xx}(\omega)$ is
\begin{equation}
\sigma^{reg}_{xx}(\omega)=\frac{\Im m[\Lambda_{xx}({\bf q}=0,\omega)]}{\omega}.
\label{eq:sigreg}
\end{equation}

The Drude weight, D$_{S}$, is the zero-frequency contribution for the real part of the spin conductivity. In our model the Drude weight is dependent on T, $\eta$, and $\Delta$ (see Fig.~\ref{fig:neeldrude}). A finite value of Drude weight indicates ballistic transport. At zero temperature D$_{S}$=0 is a signature of a spin insulator and D$_{S}>0$ of a spin conductor ~\cite{chandraJPCM7933,shastryPRL65243}. This is similar to the classification scheme for charge conductivity ~\cite{zhangPRB477995}. In the next section, Section \ref{sec:spincon}, we use the defintion of the spin current operator and the many body Green function formalism description to compute the effects of corrrelation on the spin conductivity.  

{\section{Spin conductivity}\label{sec:spincon}}
In this section we derive explicit expressions for the regular part of the spin conductivity, $\sigma_{xx}^{reg}(\omega)$. We proceed by defining the magnon propagators
\begin{eqnarray}
G_{\alpha\alpha}=-i\langle 0|\mathcal{T}\alpha_{{\bf k}}(t)\alpha^{\dag}_{{\bf k}}(0)|0\rangle,\\
G_{\beta\beta}=-i\langle 0|\mathcal{T}\beta^{\dag}_{{\bf k}}(t)\beta_{{\bf k}}(0)|0\rangle,
\end{eqnarray}
where $\mathcal{T}$ is the time ordering operator and $|0\rangle$ is the ground state wavefunction. The bare Fourier transformed propagators for the $\alpha$ and $\beta$ magnons in the absence of interactions are then given by
\begin{equation}
G^{(0)}_{\alpha\alpha}=\frac{1}{\omega -\Omega_{{\bf k}}+i0^{+}},\\
G^{(0)}_{\beta\beta}=\frac{-1}{\omega +\Omega_{{\bf k}}-i0^{+}},
\end{equation}
where $\Omega_{{\bf k}}$ is the LSWT energy dispersion. The $\alpha$ magnon propagators are represented by lines with single arrows and the $\beta$ magnon propagators are represented by lines with double arrows in the Feynman diagram. The definition of $\sigma_{xx}^{reg}(\omega)$ can now be rewritten in terms of the magnon propagators as
\begin{equation}
\frac{\sigma^{reg}_{xx}(\omega)}{(g\mu_{B})^{2}}=\frac{\Im m[\Lambda_{xx}({\bf q}=0,\omega)]}{\omega}=-\frac{\Im m[G({\bf q}=0,\omega)]}{\omega}.
\label{eq:sigreggreen}
\end{equation}
To evaluate the above we define the time-ordered spin current correlation function ~\cite{sentefPRB75214403}
\begin{equation}
G(t)=-\frac{i}{\hbar N}<0|\mathcal{T}j_{x}(t)j_{x}(0)|0>,
\label{eq:greencor}
\end{equation}
where the symbols have their usual meaning. Now utilizing the definition of j$_{x}$ from Eq.~\ref{eq:scft} and following the steps outlined in Ref ~\onlinecite{sentefPRB75214403} we can write the propagator as
\begin{eqnarray}
&G(\omega)=\frac{[2J_{x}S\alpha(S)]^{2}}{\hbar N}\sum_{\bf{k},\bf{k^{'}}}\frac{\sin {k_{x}}\sin {k_{x}^{'}}}{\varepsilon_{\bf{k}}\varepsilon_{{\bf k'}}}\Pi_{{\bf k}{\bf k'}}(\omega),
\end{eqnarray}
where $\Pi_{{\bf k}{\bf k^{'}}}(t)$ is the two-magnon propagator and is defined as
\begin{equation}
\Pi_{{\bf k}{\bf k^{'}}}(t)=-i\langle 0|\mathcal{T}\alpha_{{\bf k}}(t)\beta_{{\bf k}}(t)\alpha^{\dag}_{{\bf k'}}(0)\beta^{\dag}_{{\bf k'}}(0)|0\rangle.
\label{eq:twomag}
\end{equation}
The two-magnon propagator can be re-written in terms of the single magnon propagators and a vertex function
\begin{eqnarray}
&\Pi_{\bf{k}\bf{k}^{'}}(\omega)=i\int_{-\infty}^{\infty}\frac{d\omega'}{2\pi}G_{\alpha\alpha}(\bf{k},\omega+\omega')G_{\beta\beta}(\bf{k},\omega')\Gamma_{{\bf k}{\bf k'}}(\omega,\omega').\nonumber\\
\end{eqnarray}
The vertex function is denoted by $\Gamma_{{\bf k}{\bf k'}}(\omega,\omega')$ and satisfies the Bethe-Salpeter equation given below ~\cite{fetter}
\begin{eqnarray}
&\Gamma_{{\bf k}{\bf k'}}(\omega,\omega')=\delta_{{\bf k}{\bf k'}}-\frac{i}{\hbar}\frac{(1+\eta)\Delta J}{N}\sum_{\bf{k_{1}}}\int_{-\infty}^{\infty}\frac{d\omega_{1}}{2\pi} \mathcal{V}_{\bf{k}\bf{k_{1}}\bf{k_1}\bf{k}}^{\alpha\beta}(\omega',\omega_{1})\nonumber\\ 
&\times G_{\alpha\alpha}(\bf{k_1},\omega+\omega')G_{\beta\beta}({\bf k_1},\omega_{1})\Gamma_{{\bf k_{1}}\bf{k}}(\omega,\omega_{1}).
\label{eq:bsalp}
\end{eqnarray}

In the above expression the four-point vertex is represented by $\mathcal{V}_{\bf{k}\bf{k_{1}}\bf{k_1}\bf{k}}^{\alpha\beta}(\omega',\omega_{1})$ and contains all the irreducible interaction terms. Integral equations can be set up to solve for the expression for the spin conductivity with interactions (ladder approximation) at zero temperature. We utilize these integral equations to numerically compute the effects of spatial and spin anisotropy in the AF and CAF phase.

{\subsection{Antiferromagnetic phase}\label{subsec:scaf}}
{\subsubsection{Spin conductivity: non-interacting}\label{subsubsec:nispincon}}
We first investigate the behavior of the regular part of the spin conductivity in the zero frequency limit. This is equivalent to the replacement of the Green function with the non-interacting propagator G$^{(0)}$ and setting $\mathcal{V}_{\bf{k}\bf{k_{1}}\bf{k_1}\bf{k}}^{\alpha\beta}(\omega',\omega_{1})$=0. The magnon propagator in the non-interacting case is then written as
\begin{equation}
G^{(0)}(\omega)=\frac{[2J_{x}S\alpha(S)]^{2}}{\hbar N}\sum_{\bf{k}}\frac{\sin^{2}(k_{x})}{\varepsilon^{2}_{\bf{k}}}\frac{1}{\omega-2\Omega_{\bf{k}}+i0^{+}}.
\end{equation}
Based on the above expression we can obtain $\sigma^{reg}(\omega)$ in our non-interacting case as
\begin{eqnarray}
\sigma_{xx}^{reg}&=&\frac{(g\mu_{B})^{2}}{h}\frac{\pi^{2}}{(1+\eta)^{2}\widetilde{\omega}}\frac{2}{N}\sum_{{\bf k}}\frac{\sin^{2}(k_{x})}{\varepsilon^{2}_{{\bf k}}}\delta(\widetilde{\omega}-2\varepsilon_{{\bf k}}),\nonumber\\
\label{eq:nisigmareg}
\end{eqnarray}
where h is the Planck constant, $\widetilde{\omega}=\omega/\Omega_{max}$ with $\Omega_{max}=2J_{x}(1+\eta)S\alpha(S)/\hbar$.
\begin{figure}[t]
\centering
\includegraphics[width=3.5in]{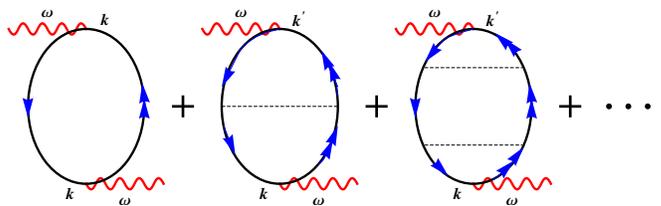}
\caption{\label{fig:ladder} (Color online) Two-magnon scattering ladder diagram contributions. The propagator with a single arrow corresponds to the $\alpha$ magnon. The propagator with two arrows correspond to $\beta$ magnons. The magnon pairs undergo repeated scattering as shown by the dashed line. The infinite sum of bubbles give rise to the interaction contribution in the spin conductivity.}
\end{figure}

In Fig.~\ref{fig:sigmani} we display the results of the regular part of the non-interacting spin conductivity, $\sigma^{reg}_{xx}(\omega)$, in units of (g$\mu_{B})^{2}$/h for the spatially anisotropic but spin isotropic ($\Delta=1$) 2D Heisenberg model in the AF phase. The plot is for $\eta=J_{y}/J_{x}$ values of 1 (black circles), 0.7 (red stars), and 0.5 (blue diamonds). The choice of anisotropy parameters and the corresponding example material system is listed in Table~\ref{tab:FCCDatatable}. From our calculations we find that decreasing $\eta$ (increasing spatial anisotropy) leads to an increase in the regular part of the spin conductivity. The finite value of the spin conductivity in the zero frequency limit is in complete contrast to the results of the spin anisotropy only model. In that model the smallest amount of spin anisotropy introduced a gap. The divergence near $\widetilde{\omega}$=2 is due to the Van-Hove singularity. This singularity is cured by introducing interactions as shown in the next section, Section~\ref{subsubsec:ladder}.

{\subsubsection{Spin conductivity: ladder approximation}\label{subsubsec:ladder}}
We obtain the corrections to the spin conductivity due to interactions within the ladder approximation of two-magnon
scattering (see Fig.~\ref{fig:ladder}). Within this approximation scheme we replace the four-point vertex, $\mathcal{V}_{\bf{k}\bf{k_{1}}\bf{k_1}\bf{k}}^{\alpha\beta}(\omega',\omega_{1})$ by the first order irreducible interaction part given by V$^{(4)}_{{\bf k}{\bf k_{1}}{\bf k_{1}}{\bf k}}$. The expression for this vertex is stated in the Appendix ~\ref{app:appvert}. Now following the procedure listed in Refs.~\onlinecite{canaliPRB457127,zeigerPRB4992} we can set up an algebraic solution of coupled integral equations based on the following decoupling schemes
\begin{eqnarray}
&\sum_{\bf{k}}\sin(k_{x}) \gamma_{\bf{k}}g_{\bf k}=0,\\
&\sum_{\bf{k}}\sin(k_{x}) \gamma_{\bf{k-k^{'}}}g_{\bf k}=\frac{\sin(k_{1x})}{1+\eta}\sum_{\bf k}\sin^{2} (k_{x})g_{\bf k}.
\label{eq:decoup}
\end{eqnarray}
Repeated applications of the above decoupling equations to the Bethe-Salpeter equation (Eq.~\ref{eq:bsalp}) and the magnon propagator leads to the following expression for the regular part of the spin conductivity in the interacting ladder approximation
\begin{eqnarray}
&\sigma_{xx}^{reg}=-\frac{(g\mu_{B})^{2}}{h}\frac{\pi}{(1+\eta)^{2}\Delta^{2}\widetilde{\omega}}\Im m[\frac{l^{(2)}-\kappa[l^{(1)}l^{(1)}-l^{(0)}l^{(2)}]}{
1+\kappa[l^{(0)}+l^{(2)}]-\kappa^{2}[l^{(1)}l^{(1)}-l^{(0)}l^{(2)}]}],\nonumber\\
&\kappa=\frac{1}{2(1+\eta)S\alpha(S)}.
\label{eq:intsigmareg}
\end{eqnarray}
To rewrite the above we utlized the definition below
\begin{eqnarray}
l^{(m)}(\widetilde{\omega})&=&\frac{2}{N}\sum_{\bf{k}}\frac{\sin^{2}(k_{x})}{\varepsilon^{2}_{\bf{k}}}\frac{1}{\widetilde{\omega}-2\varepsilon_{\bf{k}}+i0^{+}}.
\end{eqnarray}

In Fig.~\ref{fig:sigmaieta} we display the results for the regular part of the interacting spin conductivity, $\sigma^{reg}_{xx}(\omega)$, in units of (g$\mu_{B})^{2}$/h within the ladder approximation for the spatially anisotropic but spin isotropic ($\Delta=1$) 2D Heisenberg model. The plot is for $\eta=J_{y}/J_{x}$ values of  1 (black circles), 0.7 (red stars), and 0.5 (blue diamonds). Here $\widetilde{\omega}=\omega/\Omega_{max}$ where $\Omega_{max}=2J_{x}(1+\eta)S\alpha(S)/\hbar$. As in the non-interacting situation decreasing $\eta$ leads to an increase in the regular part of the spin conductivity. In contrast to the spin anisotropic Heisenberg model, the regular part does not vanish in the spatially anisotropic model in the $\widetilde{\omega}\rightarrow$ 0 limit. The Van-Hove divergence near the $\widetilde{\omega}$=2 point is cured after inclusion of interactions. Also decreasing $\eta$ leads to a shift in the peak value towards lower frequency. 
\begin{figure}[t]
\centering
\includegraphics[width=3.5in]{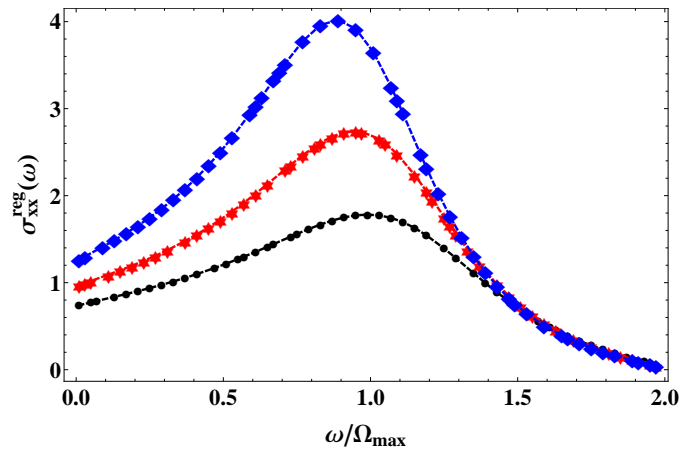}
\caption{\label{fig:sigmaieta} (Color online) Regular part of the interacting spin conductivity, $\sigma^{reg}_{xx}(\omega)$, in units of (g$\mu_{B})^{2}$/h within the ladder approximation (see Fig.~\ref{fig:ladder}) for the spatially anisotropic but spin isotropic ($\Delta=1$) 2D Heisenberg model. Here $\widetilde{\omega}=\omega/\Omega_{max}$ where $\Omega_{max}=2J_{x}(1+\eta)S\alpha(S)/\hbar$. The plot is for $\eta=J_{y}/J_{x}$ values of  1 (black circles), 0.7 (red stars), and 0.5 (blue diamonds). The dashed lines are a guide to the eye.}
\end{figure}

In Fig.~\ref{fig:sigmaietadelta} we show the results of the regular part of the spin conductivity, $\sigma^{reg}_{xx}(\omega)$, in units of (g$\mu_{B})^{2}$/h within ladder approximation with both spin and spatial anisotropy.  Here $\widetilde{\omega}=\omega/\Omega_{max}$ where $\Omega_{max}=2J_{x}\Delta(1+\eta)S\alpha(S)/\hbar$. The plot is for $\eta=0.5~\& ~\Delta=1$ (blue diamonds), $\eta=0.5  ~\& ~\Delta=1.001$ (red stars), $\eta=0.5  ~\& ~\Delta=1.01$ (green triangles), and $\eta=1.5  ~\& ~\Delta=1.01$ (black circles). Inclusion of the smallest amount of spin anisotropy suppresses the regular part of the spin conductivity and has a stronger effect than spatial anisotropy. Decreasing $\eta$ from 1.5 to 0.5 but keeping the spin anisotropy constant ($\Delta=1.01$) simply increases the magnitude of $\sigma^{reg}_{xx}(\omega)$ and leads to a shift in the peak. There is no effect on the gap above which spin conductivity is non-zero. The conductivity gap is controlled by the spin anisotropy parameter only.

The dc conductivity is given by the zero frequency limit of the regular part and is defined as, $\sigma_{xx}^{DC}=\sigma_{xx}^{reg}(\omega \rightarrow 0)$. A non-zero value of $\sigma^{DC}$ is an indicator of a non-ideal spin conductor. Within our spatially anisotropic model in the AF phase we find several parameter ranges where the system is not a spin insulator rather a non-ideal conductor. Based on the above calculations we conclude that a spatially anisotropic quantum Heisenberg AF is a better spin conductor than a spin anisotropic system at zero temperature (since the zero frequency regular spin conductivity weight survives). 

{\subsection{Drude Weight: Anisotropy and Temperature}\label{subsec:drude}}In this section we investigate the finite temperature properties of the spin Drude weight, D$_{S}(\eta,\Delta,T)$. At finite temperature AF long-range order cannot exist in the purely 2D Heisenberg model. However it is known from a non linear sigma model analysis that the spin-spin correlation length is rather long, upto to T$\sim$J/k$_{B}$, where J is the exchange constant and k$_{B}$ is the Boltzmann factor ~\cite{sudipPRL62835,dingPRB433562,nagaoJPSJ671029}. Therefore spin wave like excitations could exist in such a situation and our finite temperature analysis would still be applicable. Furthermore, anisotropy and the presence of long-range exchange interactions in real material systems can help stabilize a magnetically ordered AF state. Based on these assumptions we compute the spin Drude weight (within the collision less approximation) which is a measure of the ballistic nature of the system. Using Eq.~\ref{eq:dsweight} we obtain the following expression for the spin Drude weight, D$_{S}(\eta,\Delta,T)$
\begin{align}
\frac{D_s}{\pi}=\frac{J_{x}S\alpha(S)}{2T(1+\eta)\Delta^3}\frac{2}{N}\sum_{{\bf k}}\left(\frac{\sin(k_x)\gamma_{{\bf k}}}{\varepsilon_{{\bf k}}}\right)^{2}\frac{1}{\sinh^2\frac{\varepsilon_{{\bf k}}}{2T}}.
\label{eq:finalds}
\end{align}
For our model the spin Drude weight depends on three parameters - $\eta, \Delta$, and T. 

In Fig.~\ref{fig:neeldrude} we show the dependence of the Drude weight, D$_{S}$(T, $\eta$, $\Delta$) on temperature, spatial, and spin anisotropy. The plot is for $\eta=1~\&~\Delta=1$ (black circles), $\eta=0.7~\& ~\Delta=1$ (red squares), $\eta=0.5~\&~\Delta=1$ (blue up triangles), $\eta=1~\& ~\Delta=1.001$ (green rectangles), $\eta=1~\&~\Delta=1.01$ (orange stars), and $\eta=1~\&~\Delta=1.1$ (brown down triangles). We find that with decreasing $\eta$ (increasing spatial anisotropy) the Drude weight is enhanced at finite temperature. However increasing spin anisotropy suppresses the Drude weight at finite temperatures. In the zero temperature limit, the Drude weight vanishes for both the spatial and spin anisotropic model. But finite temperatures encourage ballistic transport due to a non-vanishing value of the drude weight. Based on the above results we conclude that a spatially anisotropic 2D quantum Heisenberg AF is a better spin conductor than a spin anisotropic AF at finite temperatures. 

The conclusions for the non vanishing Drude weight at finite temperature should be taken with a cautionary note. Within the collision less (that is non interacting) quasi particle approximation a finite Drude weight at non-zero temperatures is expected. In this approximation the magnons propagate as undamped excitations with infinite lifetime and energy ~\cite{sachdev,damlePRB568714}. However if collisions are included the magnon lifetime will become finite, the conductivity will no longer be ballistic, and will lead to a broadening of the delta function ~\cite{damlePRB568714,harrisPRB3961}. The broadening is determined by the inverse lifetime of the magnons. Although we do not perform the general analysis of the collision dominated transport ~\cite{sachdev} to treat the singular part of the spin conductivity, we believe the qualitative effects of spatial and spin anisotropy on the Drude weight as predicted from the non-interacting approximation will not be affected by the more rigorous collision dominated transport treatment. 

\begin{figure}[h]
\centering
\includegraphics[width=3.5in]{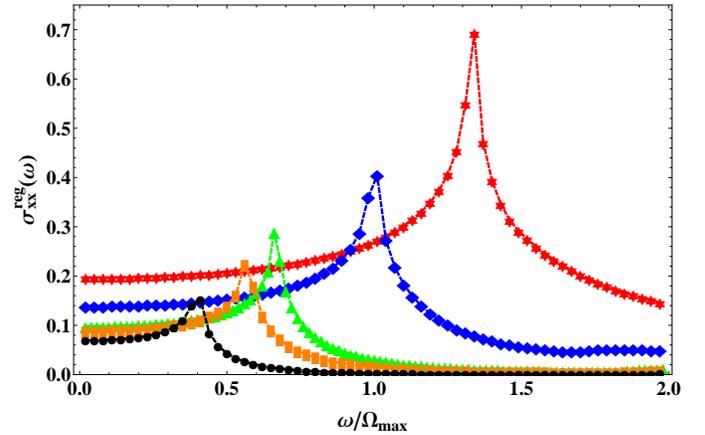}
\caption{\label{fig:caf} (Color online) Regular part of the non-interacting spin conductivity, $\sigma^{reg}_{xx}(\omega)$, in units of (g$\mu_{B})^{2}$/h, for the spatially anisotropic but spin isotropic 2D Heisenberg model in the collinear antiferromagnetic phase. The plot is for $\eta=J_{y}/J_{x}$ values of -0.25 (red stars), -0.5 (blue diamonds), and -1 (green triangles), -1.25 (orange rectangles), and -2 (black circles). The dashed lines are a guide to the eye. The regular part does not vanish in the spatially anisotropic model in the zero frequency limit for non-zero values of $\eta$.}
\end{figure}

{\subsection{Collinear antiferromagnetic phase}\label{subsec:sccaf}}
The spatially anisotropic Heisenberg model can support a CAF (stripe) phase - columnar or row. In this section we explore the effects of spatial and spin anisotropy on the regular part of the spin conductivity in the previously unexplored CAF phase. We solve the problem within the non-interacting approximation. The non-interacting limit itself presents interesting features in the spin conductivity (see Fig.~\ref{fig:caf}). To study the CAF phase we consider the following model Hamiltonian
\begin{eqnarray}
H&=&J_{x}\sum_{<i,j>}{\bf{S}}_{i}^{A}\cdot{\bf{S}}_{j}^{B}+\frac{1}{2}J_{y}\sum_{<i,j>}({\bf{S}}_{i}^{A}\cdot{\bf{S}}_{j}^{A}+{\bf{S}}_{i}^{B}\cdot{\bf{S}}_{j}^{B})\nonumber\\
&=&J_{x}\sum_{i,x}[\frac{1}{2}(S_{i}^{A+}S_{i+{\bf x}}^{B-}+S_{i}^{A+}S_{i+{\bf x}}^{B-})+\Delta S_{i}^{Az}S_{i+{\bf x}}^{Bz}]
\nonumber\\
&+&\frac{1}{2}J_{y}\sum_{i,y}[\frac{1}{2}(S_{i}^{A+}S_{i+{\bf y}}^{A-}+S_{i}^{A+}S_{i+{\bf y}}^{A-})+\Delta S_{i}^{Az}S_{i+{\bf y}}^{Az}\nonumber\\&+&\frac{1}{2}(S_{i}^{B+}S_{i+{\bf y}}^{B-}+S_{i}^{B+}S_{i+{\bf y}}^{B-})+\Delta S_{i}^{Bz}S_{i+{\bf y}}^{Bz}].
\end{eqnarray}
The superscripts A and B refer to the up- and down- spin sublattices. The exchange interaction in the x-direction is denoted by J$_{x}$ and along the y-direction by J$_{y}$. The spin vector, {\bf S}, is defined on every lattice site where the index i runs over all lattice sites and i+{\bf x} and i +{\bf y} runs over nearest neighbors in the x- and y- direction respectively. The S$^{A,B \pm}$ and S$^{A,B z}$ operators have their usual meaning. The spin anisotropy parameter is given by $\Delta$ and the spatial anisotropy parameter by $\eta=\frac{J_{y}}{J_{x}}$. As in the AF phase the Hamiltonian can be rearranged as
\begin{equation}
H=E_{0}+H_{1},
\end{equation}
where E$_{0}$ is the classical energy and H$_{1}$ is the quadratic LSWT contribution (see ~\cite{oguchi}) given by
\begin{align}
H_{1}=2J_{1x}S\sum_{\bf k}\varepsilon_{{\bf k}}\kappa_{{\bf k}}(\alpha_{\bf{k}}^{+}\alpha_{\bf{k}}+\beta_{\bf{k}}^{+}\beta_{\bf{k}}).
\end{align}
In the above we have the following
\begin{eqnarray}
&\varepsilon_{{\bf k}}=\sqrt{1-\left(\frac{\cos k_x}{\kappa_{\bf k}}\right)^2},\kappa_{{\bf k}}=\Delta-\eta(\Delta-\cos k_y),
\\ 
&\hbar\Omega_{{\bf k}}=2J_{1x}S\varepsilon_{{\bf k}}\kappa_{{\bf k}},
\\ &\hbar\Omega_{{\bf k}max}=2J_{1x}S[\Delta-\eta(\Delta+1)]=2J_{1x}SC,
\\&C=\Delta-\eta(\Delta+1).
\end{eqnarray}
Following the same procedure as in the AF phase we can define a spin current operator in the CAF phase and Fourier transform to obtain the DM representation
\begin{eqnarray}
j_{x0}&=&-2J_{1x}S\sum_{{\bf k}}\frac{\sin(k_x)}{\varepsilon_{{\bf k}}\kappa_{{\bf k}}}(\alpha_{\bf{k}}^{+}\beta_{\bf{k}}^{+}+\alpha_{\bf{k}}\beta_{\bf{k}})\nonumber\\&+&2J_{1x}S\sum_{{\bf k}}\frac{\sin(k_x)}{\varepsilon_{{\bf k}}\kappa_{{\bf k}}}\frac{B_{{\bf k}}}{A_{\bf{k}}}(\alpha_{\bf{k}}^{+}\alpha_{\bf{k}}+\beta_{\bf{k}}^{+}\beta_{\bf{k}}).
\end{eqnarray}
The coefficients A$_{\bf k}$ and B$_{\bf k}$ are given by
\begin{eqnarray}
A_{{\bf k}}=\Delta-\eta(\Delta- \cos k{_y}), B_{{\bf k}}=\cos k_{x}.
\end{eqnarray}
We then obtain the expression for the regular part of the spin conductivity in the CAF phase as
\begin{eqnarray}
\sigma_{xx}^{reg}&=&-\frac{(g\mu_{B})^2}{h}\frac{\pi}{\widetilde{\omega}C^2}\frac{2}{N}\sum_{{\bf k}}\left(\frac{\sin (k_x)}{\varepsilon_{{\bf k}}\kappa_{{\bf k}}}\right)^{2} \frac{1}{\widetilde{\omega} -2 \frac{\varepsilon_{{\bf k}}\kappa_{{\bf k}}}{C}+i0^{+}},\nonumber\\
&=&\frac{(g\mu_{B})^2}{h}\frac{\pi^2}{\widetilde{\omega}C^2}\frac{2}{N}\sum_{{\bf k}}\left(\frac{\sin (k_x)}{\varepsilon_{{\bf k}}\kappa_{{\bf k}}}\right)^{2} \delta(\widetilde{\omega} -2 \frac{\varepsilon_{{\bf k}}\kappa_{{\bf k}}}{C}).
\end{eqnarray}
where the symbols have been defined before. 

In Fig.~\ref{fig:caf} we display the results of the regular part of the interacting spin conductivity, $\sigma^{reg}_{xx}(\omega)$, in units of (g$\mu_{B})^{2}$/h, for the spatially anisotropic 2D Heisenberg model in the CAF phase. The plot is for $\eta=J_{y}/J_{x}$ values of -0.25 (red stars), -0.5 (blue diamonds), and -1 (green triangles), -1.25 (orange rectangles), and -2 (black circles). The regular part does not vanish in the zero frequency limit for non-zero values of spatial anisotropy. Contrary to the AF phase in the CAF phase increasing spatial anisotropy leads to a decrease in the spin conductivity. Also within the non-interacting approximation the spin conductivity in the CAF phase has a spike, a feature which is absent in the AF phase. The spin configuration in the CAF phase consists of ferromagnetic ordering in one direction and AF in the other. This is similar to a set of coupled 1D-AF chains in the case of a spatially anisotropic system. The sharp spike may be a signature of the one dimensional spin conductivity response ~\cite{piresPRB79064401}.

\begin{figure}[h]
\centering
\includegraphics[width=3.5in]{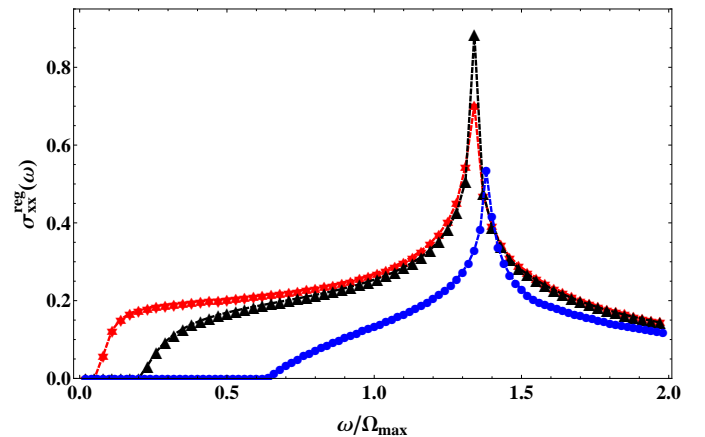}
\caption{\label{fig:cafd} (Color online) Regular part of the non-interacting spin conductivity, $\sigma^{reg}_{xx}(\omega)$, in units of (g$\mu_{B})^{2}$/h, for the spatial and spin isotropic 2D Heisenberg model in the collinear antiferromagnetic phase. The plot is for $\eta=J_{y}/J_{x}$ values of $\eta=-0.25 ~\& ~\Delta=1.001$ (red stars), $\eta=-0.25 ~\& ~\Delta=1.01$ (black triangles), and $\eta=-0.25 ~\&~ \Delta=1.1$ (blue circles). The dashed lines are a guide to the eye. Inclusion of the smallest amount of spin anisotropy $\Delta$ introduces a gap. This is similar to the spin conductivity behavior in the AF phase.}
\end{figure}

In Fig.~\ref{fig:cafd} we show the results of the regular part of the non-interacting spin conductivity, $\sigma^{reg}_{xx}(\omega)$, in units of (g$\mu_{B})^{2}$/h, for both the spatial and spin anisotropic 2D Heisenberg model in the CAF phase. The plot is for $\eta=J_{y}/J_{x}$ values of -0.25 $~\& ~\Delta$=1.001 (red stars), -0.25 $~\& ~\Delta$=1.01 (black triangles), and -0.25$~\&~ \Delta$=1.1 (blue circles). As in the AF phase, inclusion of the smallest amount of spin anisotropy suppresses the zero frequency value of $\sigma^{reg}_{xx}(\omega)$. The gap in the spin conductivity is controlled by spin anisotropy only. Finally, similar to the AF phase, we find that the spatially anisotropic CAF phase magnets are better spin conductors than those with spin anisotropy.

{\section{Summary and Conclusion}\label{sec:sum}} In this paper we analyze the effect of spatial and spin anisotropy on spin conductivity in a 2D S=1/2 Heisenberg quantum magnet on a square lattice. We explore the model in both the N\'{e}el AF phase and the CAF phase. Based on a many body Green function formalism and utilizing the Kubo formula for spin transport we explore the effects of spin and spatial anisotropy in the Heisenberg model. We utilize material parameters from SrZnVO(PO$_{4}$)$_{2}$ (spatially anisotropic) and La$_{2}$NiO$_{4}$ (spin isotropic). We find that these anisotropies have opposite effects on the magnetic system. In the AF phase spatial anisotropy does not suppress the regular part of the spin conductivity in the zero frequency limit - rather it enhances it. However spin anisotropy reduces the weight. In the CAF phase (within the non-interacting approximation) spin conductivity has a finite value for non-zero values of spatial anisotropy parameter which decreases as the spatial anisotropy is enhanced. Furthermore, we find that the CAF phase displays a spike in the spin conductivity not seen in the AF phase. The spike could be a signature of the quasi-1D behavior of the spatially anisotropic system. In both the AF and CAF phase we find that inclusion of the smallest amount of spin anisotropy causes $\sigma^{reg}(\omega)$ to develop a gap. We also explore the finite temperature Drude weight, D$_{S}(\eta, \Delta, T)$, in the AF phase for various spatial and spin anisotropy parameters. We find that within the collision less (non-interacting) quasi particle approximation the Drude weight goes to zero as the temperature approaches zero and remains non-zero at finite temperatures. Increasing spatial anisotropy increases the Drude weight value and increasing spin anisotropy decreases the Drude weight value. Based on these studies we conclude that quantum Heisenberg antiferromagnets with spatial anisotropy are better spin conductors than those with spin anisotropy at both zero and finite temperatures.  
\begin{acknowledgments}
T.D. acknowledges the invitation, kind hospitality, and research funding support from Sun Yat-sen University and Fundamental Research Funds for the Central Universities, Cottrell Research Corporation Grant (RCSA) No 20073, and Augusta State
University Katherine Reese Pamplin College of Arts and Sciences. T.D. thanks Carlo Canali, Subir Sachdev, and Philip Javernick for useful discussions. D. X. Y.  is supported by the NSFC-11074310, MOST of China 973 program (2012CB821400), Specialized Research Fund for the Doctoral Program of Higher Education (20110171110026),  Fundamental Research Funds for the Central Universities of China (11lgjc12), and NCET-11-0547.
\end{acknowledgments}
\appendix 
\section{DM Interaction Vertices\label{app:appvert}}
In this appendix we list the quartic interaction vertices obtained in the AF phase after the DM transformation of the orginal Hamiltonian, Eq.~\ref{eq:ham}. The symbols 1, 2, 3, and 4 stand for wavevectors {\bf k}$_{1}$, {\bf k}$_{2}$, {\bf k}$_{3}$, and {\bf k}$_{4}$ respectively. See Eq.~\ref{eq:uv} for the definition of u$_{{\bf k}}$, v$_{{\bf k}}$, and Eq~\ref{eq:epsil} for $\varepsilon_{{\bf k}}$. In the vertices below we use the following definitions
\begin{eqnarray}
U_{1234}=u_{1}u_{2}u_{3}u_{4},\\
x_{{\bf k}}= -\frac{v_{{\bf k}}}{u_{{\bf k}}}\left[\frac{(1-\varepsilon_{{\bf k}})}{1+\varepsilon_{{\bf k}})}\right]^{1/2},
\end{eqnarray} 
\begin{widetext}
\begin{eqnarray}
V^{(1)}&=&-U_{1234}\{x_{1}[x_{4}(x_{3}\gamma_{134}-\Delta\gamma_{14})-(\Delta x_{3}\gamma_{13}-\gamma_{1})]+x_{2}[x_{4}(x_{3}\gamma_{234}-\Delta\gamma_{24})-(\Delta x_{3}\gamma_{23}-\gamma_{2})]\},\\
V^{(2)}&=&-2U_{1234}\{x_{1}x_{2}[x_{4}(\Delta\gamma_{14}-x_{3}\gamma_{134})-(\gamma_{1}-\Delta x_{3}\gamma_{13})]+[x_{4}(\Delta\gamma_{24}-x_{3}\gamma_{234})-(\gamma_{2}-\Delta x_{3}\gamma_{23})]\},\\
V^{(3)}&=&-2U_{1234}\{x_{1}[(\Delta\gamma_{13}-x_{3}\gamma_{1})-x_{4}(\gamma_{134}-\Delta x_{3}\gamma_{14})]+x_{2}[(\Delta\gamma_{23}-x_{3}\gamma_{2})-x_{4}(\gamma_{234}-\Delta x_{3}\gamma_{24})]\},\\
V^{(4)}&=&-4U_{1234}\{x_{1}x_{2}[(x_{3}\gamma_{134}-\Delta\gamma_{14})-x_{4}(\Delta x_{3}\gamma_{13}-\gamma_{1})]+[(x_{3}\gamma_{234}-\Delta\gamma_{24})-x_{4}(\Delta x_{3}\gamma_{23}-\gamma_{2})]\},\\
V^{(5)}&=&-2U_{1234}\{x_{2}[x_{4}(\Delta x_{3}\gamma_{13}-\gamma_{1})-(x_{3}\gamma_{134}-\Delta\gamma_{14})]+x_{1}[x_{4}(\Delta x_{3}\gamma_{23}-\gamma_{2})-(x_{3}\gamma_{234}-\Delta\gamma_{24})]\},\\
V^{(6)}&=&-2U_{1234}\{x_{1}x_{2}[(\Delta x_{3}\gamma_{24}-\gamma_{234})-x_{4}(x_{3}\gamma_{2}-\Delta\gamma_{23})]+[(\Delta x_{2}\gamma_{14}-\gamma_{134})-x_{4}(x_{3}\gamma_{1}-\Delta\gamma_{13})]\},\\
V^{(7)}&=&-U_{1234}\{x_{1}[(\gamma_{134}-\Delta x_{3}\gamma_{14})-x_{4}(\Delta\gamma_{13}-x_{3}\gamma_{1})]+x_{2}[(\gamma_{234}-\Delta x_{3}\gamma_{24})-x_{4}(\Delta\gamma_{23}-x_{3}\gamma_{2})]\},\\
V^{(8)}&=&-U_{1234}\{x_{2}[x_{4}(x_{3}\gamma_{134}-\Delta\gamma_{14})-(\Delta x_{3}\gamma_{13}-\gamma_{1})]+x_{1}[x_{4}(x_{3}\gamma_{234}-\Delta\gamma_{24})-(\Delta x_{3}\gamma_{23})-\gamma_{2}]\},\\
V^{(9)}&=&-U_{1234}\{x_{2}[(\gamma_{134}-\Delta x_{3}\gamma_{14})-x_{4}(\Delta\gamma_{13}-x_{3}\gamma_{1})]+x_{1}[(\gamma_{234}-\Delta x_{3}\gamma_{24})-(x_{4}(\Delta\gamma_{23}-x_{3}\gamma_{2})]\}.
\label{eq:vertexexp}
\end{eqnarray}
\end{widetext}
\bibliography{scref}

\begin{thebibliography}{43}
\bibitem [{ogu()}]{oguchi}%
  \BibitemOpen
  \href@noop {} {}\bibinfo {note} {Inclusion of Oguchi corrections do not
  change the overall qualitative behavior of the non-interacting spin
  conductivity results in the collinear antiferromagnetic phase. We therefore
  ignore this correction in our calculation.}\BibitemShut {Stop}%
\end{thebibliography}
\end{document}